\documentclass{article} 
\usepackage{iclr2026_conference,times}
\usepackage{listings}
\usepackage{xcolor}
\usepackage{graphicx}
\usepackage{fancyhdr}
\usepackage{subcaption}

\definecolor{codegray}{rgb}{0.5,0.5,0.5}
\definecolor{codeblue}{rgb}{0.0,0.0,0.6}
\definecolor{codered}{rgb}{0.6,0.0,0.0}
\definecolor{codegreen}{rgb}{0.0,0.5,0.0}
\definecolor{codebg}{rgb}{0.96,0.97,0.98}

\lstdefinestyle{python}{
  language=Python,
  basicstyle=\ttfamily\footnotesize,
  keywordstyle=\color{codeblue}\bfseries,
  stringstyle=\color{codegreen},
  commentstyle=\color{codegray}\itshape,
  breaklines=true,
  frame=single,
  backgroundcolor=\color{codebg},
  numbers=left,
  numberstyle=\tiny\color{codegray},
  showstringspaces=false,
  columns=fullflexible
}

\lstdefinelanguage{diff}{
  morecomment=[f][\color{codegreen}]{+},
  morecomment=[f][\color{codered}]{-},
  morecomment=[f][\color{codegray}]{@},
}

\lstdefinestyle{diff}{
  language=diff,
  basicstyle=\ttfamily\footnotesize,
  breaklines=true,
  frame=single,
  backgroundcolor=\color{codebg},
  numbers=left,
  numberstyle=\tiny\color{codegray},
  showstringspaces=false,
  columns=fullflexible
}

\usepackage{amsmath,amsfonts,bm}

\def\eqref#1{equation~\ref{#1}}

\def\1{\bm{1}}

\DeclareMathAlphabet{\mathsfit}{\encodingdefault}{\sfdefault}{m}{sl}
\SetMathAlphabet{\mathsfit}{bold}{\encodingdefault}{\sfdefault}{bx}{n}

\usepackage{hyperref}
\usepackage{url}

\title{The Limits of Long-Context Reasoning in Automated Bug Fixing}

\author{Ravi Raju\thanks{Equal contribution.}, Mengmeng Ji\footnotemark[1], Shubhangi Upasani, Bo Li \& Urmish Thakker \\
SambaNova Systems\\
San Jose, CA 95134, USA \\
\texttt{\{ravi.raju,mengmeng.ji\}@sambanovasystems.edu} \\
}

\iclrfinalcopy 

\usepackage{fancyhdr}

\begin{document}
\maketitle
\begin{abstract}

Rapidly increasing context lengths have led to the assumption that large language models (LLMs) can directly reason over entire codebases. Concurrently, recent advances in LLMs have enabled strong performance on software engineering benchmarks, particularly when paired with agentic workflows. In this work, we systematically evaluate whether current LLMs can reliably perform long-context code debugging and patch generation. Using SWE-bench Verified as a controlled experimental setting, we first evaluate state-of-the-art models within an agentic harness (mini-SWE-agent), where performance improves substantially: GPT-5-nano achieves up to a 31\% resolve rate on 100 samples, and open-source models such as Deepseek-R1-0528 obtain competitive results. However, token-level analysis shows that successful agentic trajectories typically remain under 20k-30k tokens, and that longer accumulated contexts correlate with lower success rates, indicating that agentic success primarily arises from task decomposition into short-context steps rather than effective long-context reasoning. To directly test long-context capability, we construct a data pipeline where we artificially inflate the context length of the input by placing the relevant files into the context (ensuring perfect retrieval recall); we then study single-shot patch generation under genuinely long contexts (64k tokens). Despite this setup, performance degrades sharply: Qwen3-Coder-30B-A3B achieves only a 7\% resolve rate at 64k context, while GPT-5-nano solves none of the tasks. Qualitative analysis reveals systematic failure modes, including hallucinated diffs, incorrect file targets, and malformed patch headers. Overall, our findings highlight a significant gap between nominal context length and usable context capacity in current LLMs, and suggest that existing agentic coding benchmarks do not meaningfully evaluate long-context reasoning.
\end{abstract}
\section{Introduction}
Supported context lengths in large language models (LLMs) have expanded dramatically, fostering the expectation that models can directly reason over entire codebases in a single pass \citep{deepseekai2025deepseekr1incentivizingreasoningcapability, qwen3technicalreport}. Meanwhile, recent advances in LLMs have led to strong performance on software engineering tasks such as code generation and debugging \citep{jimenez2024swebenchlanguagemodelsresolve,liu2023repobenchbenchmarkingrepositorylevelcode,li2025longcodeubenchmarkinglongcontextlanguage}. Whether current LLMs can effectively use such long contexts in realistic programming scenarios, however, remains unclear.

In this work, we study the limits of long-context code reasoning by evaluating LLMs on repository-scale debugging tasks using SWE-bench Verified as a controlled experimental setting. SWE‑bench Verified is a curated subset of the SWE-bench benchmark consisting of real GitHub issues paired with code repositories and human-validated patches, where each problem has been carefully verified to ensure the issue description, tests, and environment reliably reproduce the bug. We begin by examining conventional ways to evaluate LLMs' agentic coding capabilities through open-source agent scaffolding such as SWE-agent and OpenHands \citep{wang2025openhandsopenplatformai}. We test current SOTA models like GPT-5-nano, Deepseek R1-0528 and Qwen3-32B, for their accuracy on SWE-bench Verified and how many tokens, on average, are produced in order to complete the aforementioned task. 

Reasoning models like Deepseek R1-0528 and Qwen3-32B are known to produce a significant amount of output tokens for a given task. Therefore, on first glance, one may assume that a software engineering task like SWE-bench, which may require multiple invocations of the agentic system, to solve will invariably end up being long context. However, we find that through token length analysis of agentic trajectories that they tend to be under 20k-30k tokens, clearly not long context by current standards. This demonstrates that strong agentic performance should not be interpreted as evidence of long-context capability.

In order to isolate the impact of the LLM's coding ability and to isolate long context coding performance, we examine single-shot patch generation under long contexts (64k), where all files required to construct the ground-truth patch are explicitly provided, eliminating retrieval as a confounding factor. Under this setup, state-of-the-art models, such as Qwen3-Coder-30B-A3B and GPT-5-nano, perform poorly: at 64k context, resolve rates remain in the single digits or zero (much lower than the agentic resolve rates). These failures include manifesting as hallucinated edits which result in  malformed diffs that cannot be applied as well as referencing files which don't exist in the context. 

These findings indicate that increased context length alone does not translate into reliable long-context reasoning. In contrast, the same models achieve substantially higher resolve rates when evaluated in agentic frameworks (\textit{mini-swe-agent}) that decompose the task into multiple short-context steps. While effective, such workflows largely bypass long-context reasoning rather than exercising it.

Our contributions are as follows:
\begin{enumerate}
    \item We provide token-length analysis of reasoning traces to demonstrate that common agentic coding evaluations do not meaningfully exercise long-context reasoning.
    \item We present qualitative observations of SOTA LLMs failure modes in long-context, single-shot code patch generation.
\end{enumerate}

Together, these findings highlight a substantial gap between nominal context length and usable context capacity in current LLMs for software engineering, and motivate the need for models that explicitly target long-context reasoning rather than assuming it emerges implicitly from agentic interaction.
\begin{figure}[t]
\centering
\includegraphics[width=0.8\linewidth]{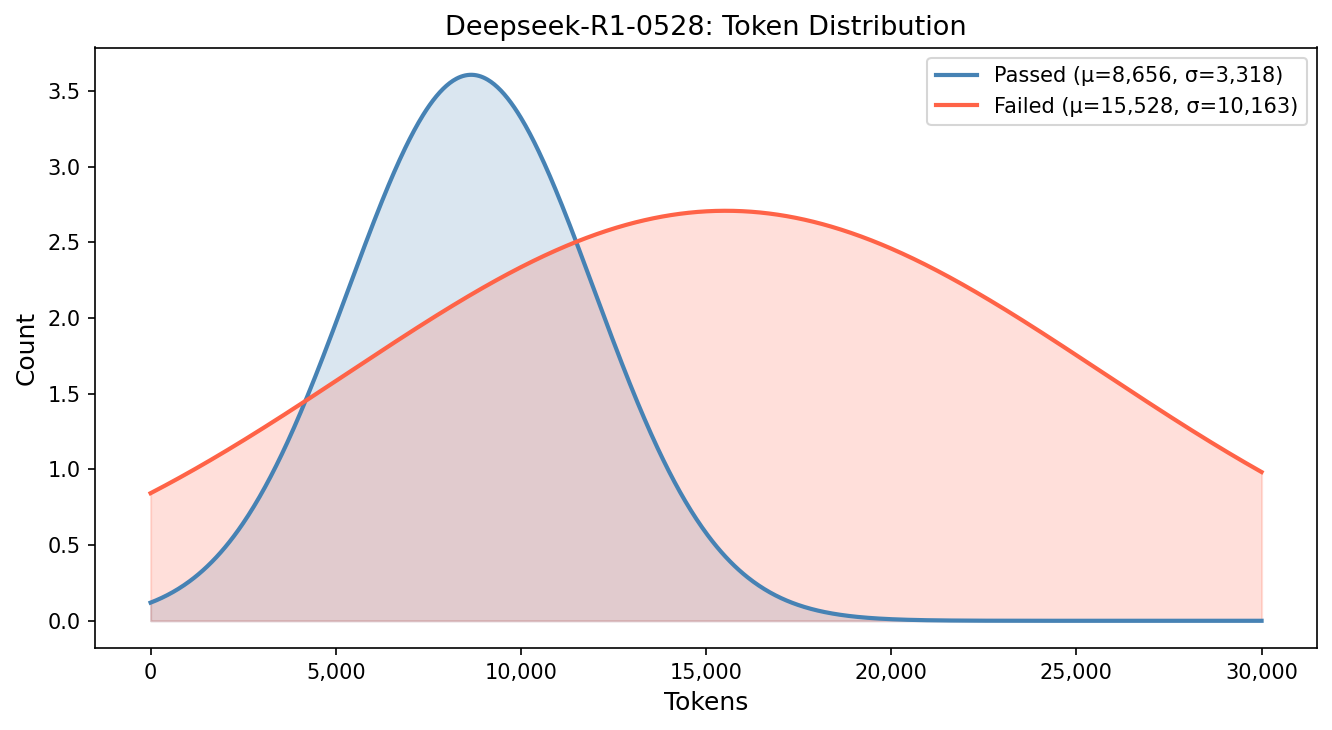}\\[2pt]
\includegraphics[width=0.8\linewidth]{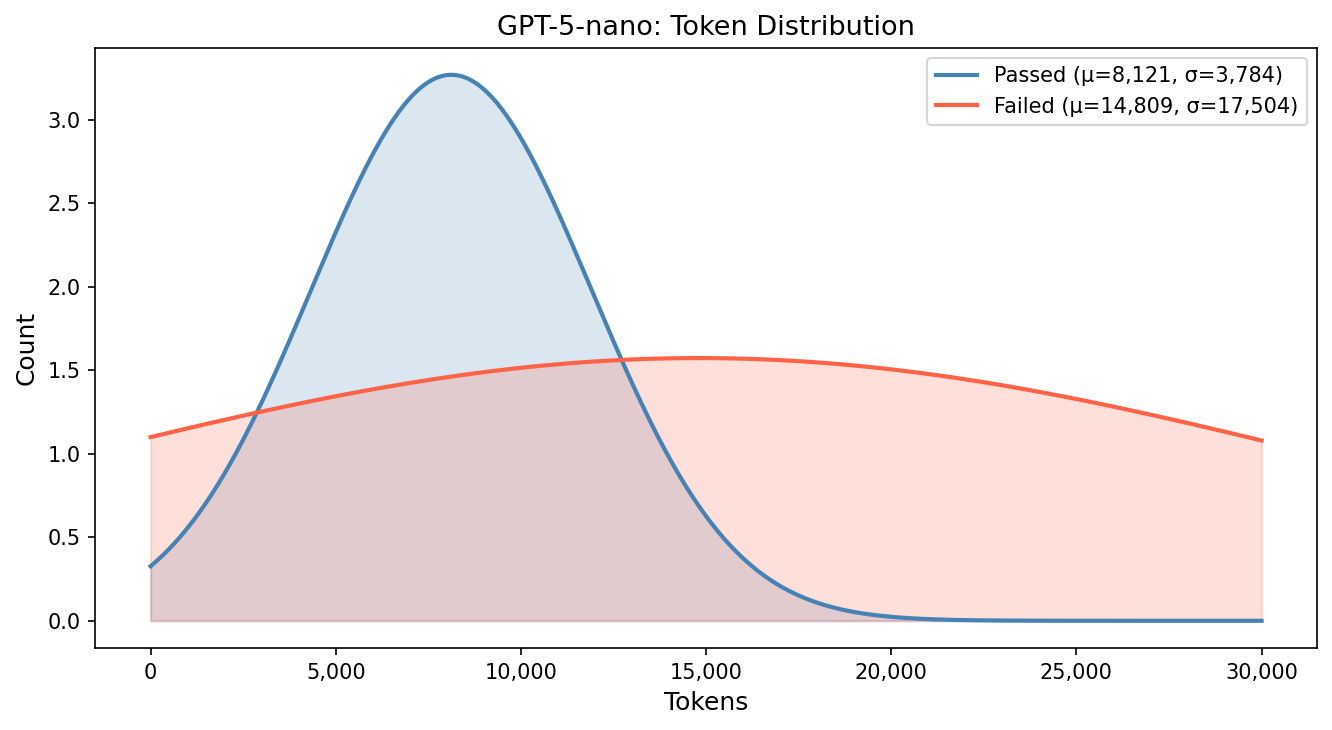}
\caption{We show the token distributions of the model trajectories for Deepseek-R1-0528 and GPT-5-nano in terms of mean and variance for passed/failed category. Deepseek-R1-0528 has 30 samples and 69 samples in the passed and fail category, respectively, whereas GPT-5-nano has 31 samples and 69 samples in the passed and fail category. The distribution for each model shows that samples which the model fail to answer tend to consume more tokens. These figure illustrate that the average conversation/trajectory is still under 20-30k and can't be considered a long context interaction.}
\label{fig:combined_sub}
\end{figure}

\section{Comparative Analysis of Agentic and Long-Context Reasoning}
We begin by evaluating LLMs on agentic workflows based on \textit{mini-swe-agent}, analyzing how task decomposition alters effective context usage \cite{yang2024sweagent}. We then contrast these results with single-shot patch generation under genuinely long contexts using retrieval-augmented inputs with perfect recall, exposing the failure modes of direct long-context reasoning. We provide more detailed descriptions of the benchmark in the Appendix \ref{SWE_description}.

\subsection{Evaluating Usable Context in Agentic Frameworks}
We evaluate SWE-bench in an agentic workflow setting using the \textit{mini-SWE-agent} framework, the reference agent released by the SWE-bench authors \citep{yang2024sweagent}. Unlike SWE-agent, which leverages an ACI (Agent-Computer Interface), mini-SWE-agent is a bash-only command-line workflow that provides a lightweight sandbox for experimenting with different models as shown in Figure \ref{fig:mini-swe-bench} (in Appendix). It produces a linear history in which each agent step is appended directly to the message stream—making it especially useful for debugging, fine-tuning, and measuring token usage and conversation length. This design aligns well with our goal of assessing SWE-bench’s strengths as an agentic benchmark while also probing its potential as a long-context benchmark.  

SWE-bench has multiple different version: SWE-bench, SWE-bench Lite and SWE-bench Verified. SWE-bench Verified is the best out of all these options as it removes unsolvable or irrelevant issues and categorizing problems by difficulty \citep{chowdhury2024swebenchverified}. As a sanity check for our evaluation infrastructure, we tested GPT-5-nano and obtained a 31\% resolve rate on 100 samples from SWE-bench Verified using the \textit{sb-cli} tool, the official SWE-bench command-line evaluation tool. GPT-5 nano's results on the open-source leaderboard on the full dataset is 34.8\%, which both derisks our evaluation infrastructure and justifies our choice of using a subsample of the dataset for evaluation \citep{chowdhury2024swebenchverified}. With our harness's fidelity verified, we evaluate 2 SOTA open-source models, DeepSeek R1-0528 and Qwen3-32B using the SambaNova Cloud API \citep{deepseekai2025deepseekr1incentivizingreasoningcapability,qwen3technicalreport}. On 100 samples from SWE-bench Verified, DeepSeek R1-0528 achieved a 30.3\% resolve rate, while Qwen3-32B reached 15.2\%. 

While these experiments confirm that SWE-bench is well-suited for agentic workflows, they also raise the question: \textit{can it still serve as a long-context benchmark}? A key challenge lies in measuring and controlling context length in an agentic setting. Context length here depends on the number of dialogue rounds and accumulated tokens. In Figure \ref{fig:combined_sub}, we show the trajectory token distributions for both Deepseek-R1-0528 and GPT-5-nano (please refer to the Appendix for Qwen3-32B's token distribution) full conversations generally stayed under 20k-30k tokens. Within this range, successful solutions tended to use shorter contexts, while samples with longer contexts tended to have lower resolve rates. It is unclear whether or not if longer contexts have worse resolve rates due to the fact that more context could confuse/complicate reasoning vs. more challenging issues require reasoning and are harder to resolve; disentangling these two factors is a good direction for future work. In totality, these findings indicate that SWE-bench is most suitable as an agentic workflow benchmark, but is less effective for evaluating long-context coding capabilities.

\subsection{Stress-Testing Direct Reasoning in 64k Context Windows}
In order to test solely the long context coding ability of models, we need to create long context evaluation with the existing tools. However, unlike SWE-bench Lite, SWE-bench Verified provides no oracle sources or files. As a result, we explored RAG techniques to retrieve long context(s) for model generation. 

\begin{figure}
    \centering
    \includegraphics[width=0.85\linewidth]{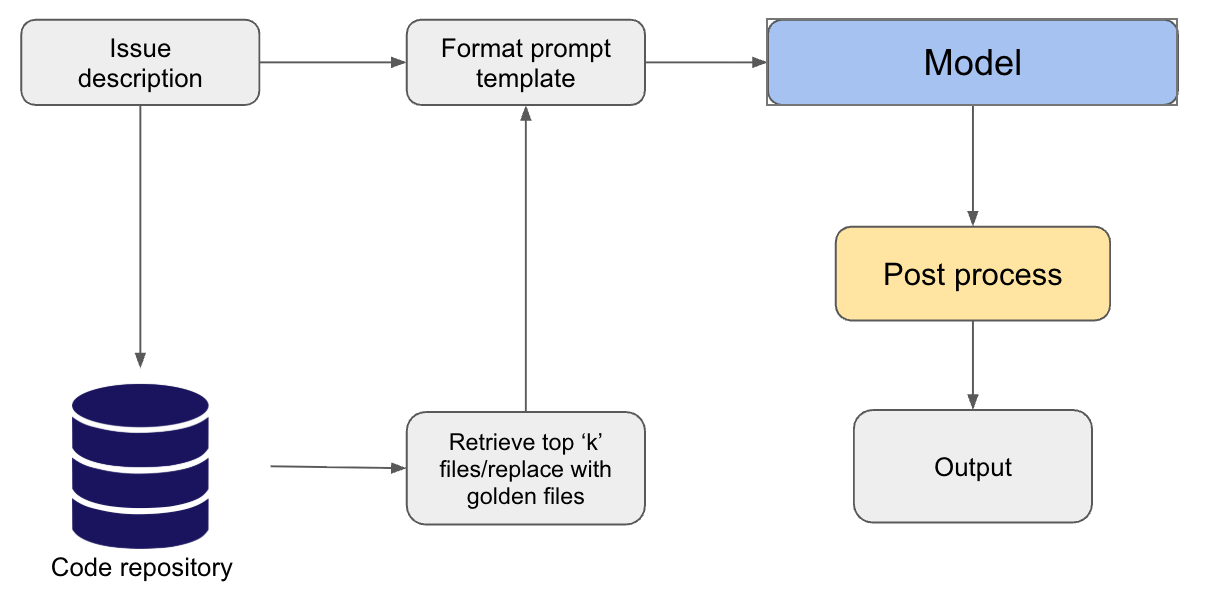}
    \caption{Our long-context gold patch generation pipeline. By retrieving/injecting all the files present in the ground truth gold patch, we isolate the LLM's patch generation ability.}
    \label{fig:long_patch_gen}
\end{figure}

\begin{lstlisting}[style=diff, caption={A misaligned patch generated by GPT-5 nano. One of the errors can be seen in the chunk header, where the line numbers from the first line clearly exceed the patch length.}, label={lst:misaligned_patch}]
--- a/lib/matplotlib/offsetbox.py
+++ b/lib/matplotlib/offsetbox.py
@@ -159,26 +159,37 @@ def _get_aligned_offsets(hd_list, height, align="baseline"):
-    if height is None:
-        height = max(h for h, d in hd_list)
-    _api.check_in_list(
-        ["baseline", "left", "top", "right", "bottom", "center"], align=align)
- 
-    if align == "baseline":
-        height_descent = max(h - d for h, d in hd_list)
-        descent = max(d for h, d in hd_list)
-        height = height_descent + descent
-        offsets = [0. for h, d in hd_list]
-    elif align in ["left", "top"]:
-        descent = 0.
-        offsets = [d for h, d in hd_list]
-    elif align in ["right", "bottom"]:
-        descent = 0.
-        offsets = [height - h + d for h, d in hd_list]
-    elif align == "center":
\end{lstlisting}

Based on our literature survey, major retrieval methods are either based on sparse retrieval methods like BM25, or more complex agentic systems, like meta-RAG, Code-RAG bench, or a hybrid variant like CodeMonkeys \citep{tawosi2025metaraglargecodebasesusing, wang2025coderagbenchretrievalaugmentcode,ehrlich2025codemonkeysscalingtesttimecompute, robertsonbm25}. Given the high cost of agent-based retrieval, we opted for a simpler yet effective strategy: use BM25 to rank code chunks and then inject the “gold” patches to guarantee high recall. With this approach, we curated a smaller dataset from SWE-bench Verified (using the same dataset as above), containing 100 samples with 100\% recall (meaning all the required files are within the context), one with 64k token context length. We illustrate our pipeline in Figure \ref{fig:long_patch_gen}.

Using this long-context dataset, we evaluated two models: GPT-5-nano and Qwen3-Coder-30B-A3B, with \textit{sb-cli} \cite{qwen3technicalreport,singh2025openaigpt5card}. We chose to evaluate these models instead of Deepseek-R1-05-28 or Qwen3-32B since at the time of writing, we didn't have access to the compute resources which would allow us to run inference at the long sequence length we desired.

On our 64k setting, the results were poor: Qwen3-Coder-30B-A3B achieved only a 7\% resolve rate, while GPT-5-nano solved none of the tasks. We conducted qualitative analysis on the model’s outputs and  revealed common failure modes. Many generated patches contained hallucinated information: some had incorrect line numbers in the diff header as shown in Listing \ref{lst:misaligned_patch}, while others targeted files that did not exist in the repository. After manually verifying that our post-processing pipeline was not the source of these errors, we conclude that single-shot prompting under genuinely long contexts exceeds the effective reasoning capacity of current LLMs for repository-scale debugging. While these models nominally support long context windows, our results indicate that, when faced with complex, multi-file code changes, they frequently fail to reliably synthesize and act on dispersed information, leading to hallucinated or malformed patches. These findings may suggest that using specialized harnesses are critical to achieving competitive performance on SWE tasks, as seen in \cite{erdil2025swebenchskills}. As a result, SWE-bench is not suitable as a single-shot long-context standalone benchmark, since failures persist even under perfect retrieval and instead reflect practical limits in current LLMs’ ability to operate at this scale.
\section{Conclusion}
Although SWE-bench is valuable for testing coding agents, it is not suitable for long-context evaluation. Our results challenge the assumption that repeated calls in agentic settings naturally lead to long-context use. In RAG/oracle experiments, models failed to solve long-context tasks in a single shot. Recently, LongCodeBench introduced LongSWE-bench, a similar benchmark to our RAG/oracle setting, showing that open-source models achieved only single-digit solve rates, and the best closed-source model (Gemini 2.5 Pro) solved just 22\% \cite{rando2025longcodebenchevaluatingcodingllms}. In essence, our observations are two-fold: 1) we need more long-context focused versions of SWE-bench to test long horizon coding tasks and 2) the current model capabilities are not sufficient to perform well on long-context (single-shot) code debugging. Taken together, these results underscore the need for targeted advances in software engineering agents that explicitly address long-context reasoning, rather than assuming such capabilities will emerge naturally from agentic interaction alone.
\section{Ethics Statement}

This work focuses on evaluating the long-context capabilities of existing large language models using established, publicly available benchmarks and evaluation frameworks. All experiments are conducted on open-source datasets (SWE-bench and SWE-bench Verified) derived from publicly accessible GitHub repositories, and we do not introduce new data collection involving human subjects or sensitive personal information.

Our study does not involve training or deploying models in real-world software development environments. Instead, it analyzes model behavior in controlled offline settings to better understand the limitations of long-context reasoning and agentic workflows. While our findings highlight failure modes such as hallucinated or malformed patches, these results are presented for diagnostic and benchmarking purposes and are not intended to encourage automated code deployment without human oversight.

We believe that clarifying the limits of current LLM capabilities, particularly in safety-critical tasks such as code modification, is a positive contribution that can help prevent misuse and overreliance on automated systems. No foreseeable ethical risks arise beyond those already associated with the use of large language models for code generation, which are mitigated by the evaluation-only nature of this work.

\bibliography{iclr2026_conference}
\bibliographystyle{iclr2026_conference}
\appendix
\section{Appendix}

\subsection{SWE-bench benchmark description}
\label{SWE_description}
SWE-bench is a benchmark designed to evaluate large language models on complex real-world software engineering tasks drawn from GitHub \citep{jimenez2024swebenchlanguagemodelsresolve}. Given a repository and a GitHub issue, a model must produce a patch that resolves the problem described. Each instance in the dataset consists of a failing test (Fail-to-Pass) that is fixed by the corresponding pull request along with additional tests to ensure no unrelated behavior is broken (Pass-to-Pass). Each example in the benchmark contains a "gold" patch file, which has the code which are able to resolve the problem mentioned in the issue description and pass both suites of tests. The tasks come from actual GitHub issues and their associated fixes, making the evaluation grounded in how software evolves in practice. While many recent works focus on SWE-bench in the context of agentic / file-retrieval workflows (given its complexity), we set out to explore its suitability as a long-context benchmark, considering how it uses retrieved context from the entire repositories as the context.

The SWE-bench evaluation pipeline has two main stages: \textbf{generation} and \textbf{evaluation}. In the generation stage, the model is prompted with an issue description and relevant repository context, and it produces a candidate patch. In the evaluation stage, this patch is tested inside a controlled Docker environment: 1. the repository is cloned, 2. the model’s patch is applied to the targeted file(s), and 3. the full test suite is run to check whether the issue is resolved without breaking other functionality.

\subsection{mini-swe-agent Diagram}
\begin{figure}
    \centering
    \includegraphics[width=0.9\linewidth]{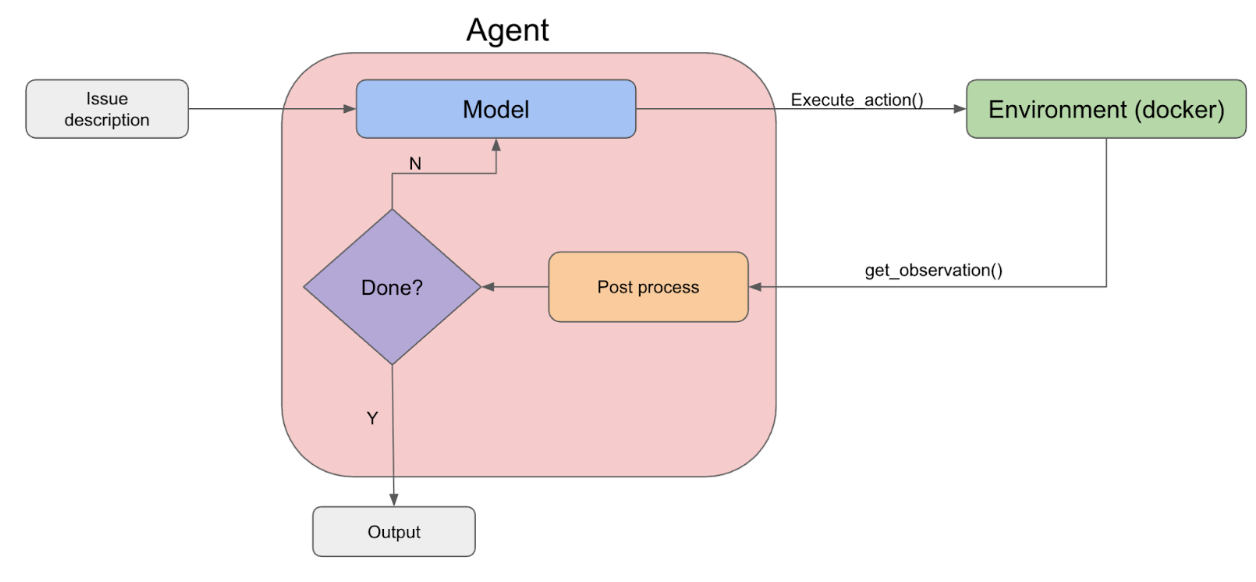}
    \caption{A block diagram of the mini-swe-agent framework.}
    \label{fig:mini-swe-bench}
\end{figure}

In Figure \ref{fig:mini-swe-bench}, we show a basic flowchart of the control flow of the agentic harness.

\subsection{Qwen3-32B Token Distribution}
\begin{figure}
    \centering
    \includegraphics[width=0.9\linewidth]{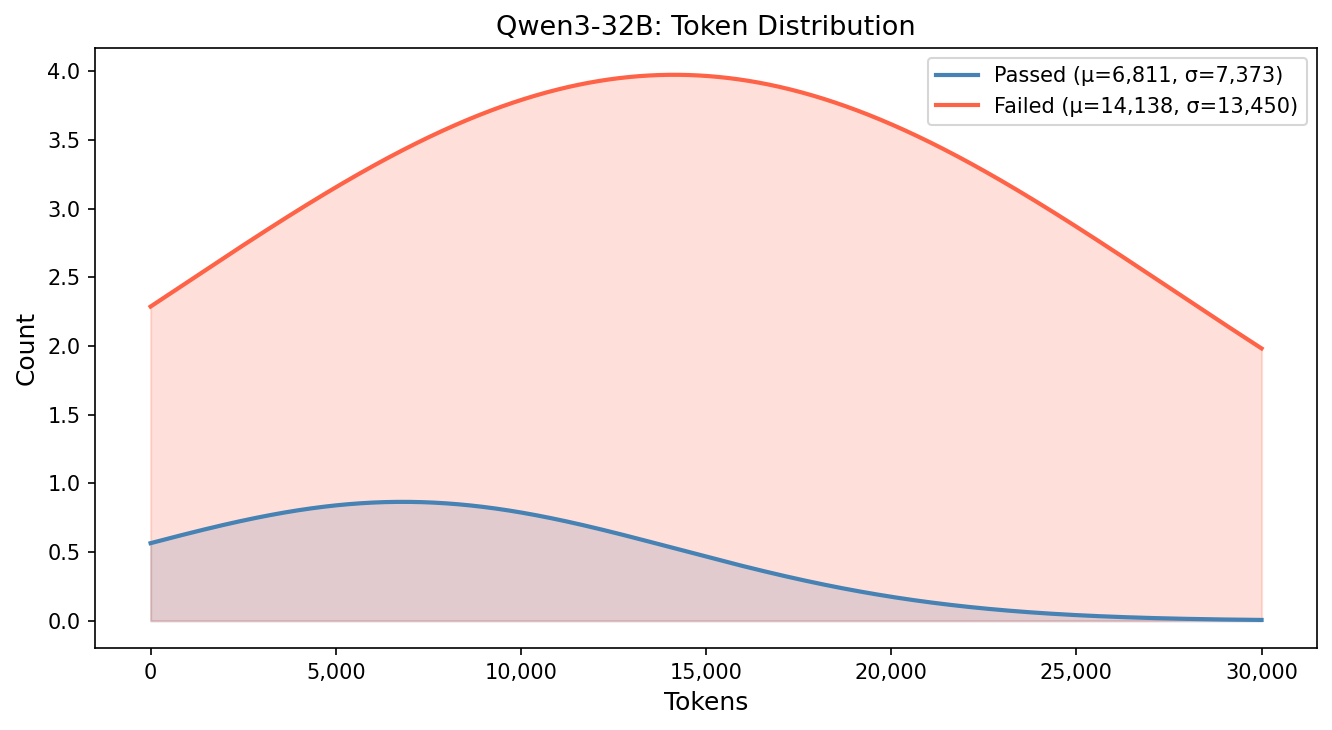}
    \caption{Token distribution of Qwen3-32B. The passed category includes 16 samples whereas the failed category includes 88 samples.}
    \label{fig:qwen3-32b token distribution}
\end{figure}

In figure \ref{fig:qwen3-32b token distribution}, we show the token distribution of the traces with Qwen3-32B on both the passed and failed categories. One difference that we noted compared to Deepseek-R1-0528 and GPT-5-nano is that we had many more samples which seemed to have failed so we increased the sample set from 100 to 150 and ended up with 104 valid runs. Ultimately, we still observe a similar trend that failed trajectories consume more tokens than successful ones and that the vast majority are under 20-30k tokens.

\end{document}